%Paper: solv-int/9407001
%From: asher@venus.fiz.huji.ac.il (Asher Yahalom )
%Date: Fri, 1 Jul 94 14:38:19 +0300

\magnification=\magstep1
\def\refs{\leftskip=.3truein\parindent=-.3truein}
\def\endrefs{\leftskip=-.3truein\parindent=.3truein}
\baselineskip=24pt

\centerline{HELICITY CONSERVATION VIA THE NOETHER THEOREM}

\bigskip

\centerline{Asher Yahalom}

\centerline{The Racah Institue of Physics, Jerusalem 91904, Israel}
\bigskip

\centerline{\bf Abstract}
\noindent
The conservation of helicity in ideal barotropic fluids is discussed from a
group
 theoretical point of view. A new symmetry group is introduced  i.e. the alpha
group of
 translations. It is proven via the Noether theorem that this group generates
helicity
 conservation.

\bigskip


\bigskip

\noindent
{\bf 1. Introduction}
\bigskip

The conservation of helicity $ {{\cal H}} = \int \vec u \cdot \vec \omega  d^3x
$ in
ideal barotropic fluid when certain conditions are satisfied in
particular, when vorticity lines  are closed was discovered by Moffat (1969).
 Moreau (1977) has discussed the conservation of helicity from
the group theoretical point of view. In his paper he used
an enlarged  Arnold  symmetry group (Arnold(1966)) of fluid element labeling to
generate the conservation of helicity. Lynden-Bell and Katz (LBK 1981)
introduced a
special labeling
   with the help of the concepts "load" and "metage" defined by those
authors .
 The purpose of this paper is to show that the symmetry group
generating conservation of helicity becomes a very simple one parameter
translation
group in the space of labels (alpha space) when represented by LBK labeling.

\vfill\eject

\noindent
{\bf 2.  Noether theorem and the conservation of Helicity}
\bigskip

	It will be useful to write the equations of motion
 first to show our essentially standard notations:
${\vec r} =(x^k)= (x,y,z) ~~[k,l,m,n = 1,2,3]$ for the
 position of fluid elements labeled $(\alpha^k)$, $t$ for time,
 ${\vec u \equiv {D\vec r \over Dt}}$ for the velocity field in
 inertial coordinates, $\rho$  the  density, $P(\rho)$
 the pressure, $h(\rho) = \int dP/\rho$ the specific enthalpy, $ \varepsilon $
is the
specific internal energy, $\Phi $ is an external potential.
Euler's equations for barotropic flows may be written as:
$$ \vec {\cal O} \equiv {D\vec u \over Dt} +
    \vec \nabla (h + \Phi) = 0,  \eqno(2.1)$$
The vorticity is defined by:
$$ \vec \omega = {\rm rot}~ \vec u         \eqno(2.2)$$
Now let us denote the initial position of a fluid element by $ (x^k_0) $,
 by mass conservation:
$$ \rho(x^k)d^3x = \rho(x^k_0) d^3x_0 = \rho(x^k_0){\partial (x^1_0, x^2_0,
x^3_0)
\over \partial (\alpha^1,\alpha^2,\alpha^3)}d^3\alpha \eqno(2.3)$$
Since the initial position of a fluid element can not depend on time it must
depend on
 the label only, and therefore by an appropriate choice of the $ \alpha 's $ we
obtain:
$$ \rho(x^k)d^3x = d^3\alpha , \qquad \rho={\partial (\alpha^1, \alpha^2,
\alpha^3)
 \over \partial (x^1, x^2, x^3)} \eqno(2.4) $$
Where we assume that the above expressions of course exist.
The action (Bretherton 1970) is given by:
$$ A = \int ^{t_1}_{t_0} L dt \quad with \quad L = \int _{V} \left[{1 \over 2}
   \vec u^{2} - \left(\varepsilon(\rho) +  \Phi(\vec r) \right)
   \right] \rho d^{3}x = \int  \left[{1 \over 2}
   \vec u^{2} - \left(\varepsilon(\rho) + \Phi(\vec r)\right)
   \right] d^{3}\alpha \eqno(2.5)$$
Following Bretherton we  obtain the variation of action:
$$ \Delta A =  \int \int [{D \vec u \cdot \vec \xi \over Dt} - \vec {\cal O}]
\cdot \vec
 \xi d^3\alpha dt = \int \vec u \cdot \vec \xi  d^3\alpha  \bigg|_{t_0}^{t_1} -
 \int \int \vec {\cal O} \cdot \vec \xi d^3\alpha dt.  \eqno(2.6) $$
Where $\vec \xi(\vec \alpha,t)$ is an arbitary small displacement of a fluid
element labeled
$\vec \alpha$. Now if variations  disappear at times ${t_0},{t_1}$ than A is
extremal only if  Euler's equations are satisfied. If on the other hand we
make a symmetry displacement i.e. a displacement that makes $\Delta A$ vanish
and assume that Euler's equations are satisfied, we obtain that:
$$ \int_V \vec u \cdot \vec \xi  d^3\alpha = const. \eqno(2.7) $$
this is Noether Theorem in it's fluid mechanical form. A similiar formalism can
be
constructed for liquids ($\rho = const.$) by replacing $ \varepsilon
\rightarrow h = {P
\over \rho}$ in (2.5), equation (2.7) is not altered.

The $\alpha's$ so chosen as to satisfy the equation
(2.4)  are not unique in fact one can always choose another set of variables
say
$\tilde \alpha's$ such that:
$$ {\partial (\tilde\alpha^1,\tilde\alpha^2,\tilde\alpha^3) \over
 \partial (\alpha^1,\alpha^2,\alpha^3)} = 1. \eqno(2.8)  $$
It is quite clear that if the domain of integration is not modified any new set
of
 $\alpha's$
satisfying (2.8) can be chosen without effecting the value of the Lagrangian
$L$.
This is nothing but Arnolds (1966)
alpha space symmetry group under which $L$ is invariant. For some flows the
domain of
 integration can be modified
 without effecting $L$, in that case we have additional elements in our
symmetry group.
 If we make only small changes $\delta \alpha$ than we can define the group as
follows:
$$ {\partial \delta \alpha_k \over \partial \alpha_k} = 0 \eqno(2.9.a)$$
$$\delta\vec\alpha\cdot \vec n\bigg|_{surface} = 0 \eqno(2.9.b)$$
where $\vec n$ is a unit vector orthogonal to the surface of the alpha space
volume
 which we integrate over. The restriction (2.9.b) is only needed when the
infinitesimal
 transformation changes the domain of integration in such a way as to modify
$L$.
 In this paper we are interested in the subgroup of translation, i.e.:
$$ \delta \alpha_k =a_k \qquad a_k =const. \eqno(2.10)$$
 This subgroup of course  does not satisfy (2.9.b) unless
 at least few of the $\alpha's$ are cyclic or $L$ is not effected by the
modification
 of domain.

 LBK(1981) have defined the three  following  parameters: the load $\lambda$,
 the metage $\mu$ and  $\beta$. Surfaces of constant load $\lambda$ are
surfaces of
constant mass per unit
vortex strength in a narrow tube of vortex lines: if $dM$ is the mass in a tube
with vorticity flux $dC$,
$$ \lambda = {d M \over d C} = \int_{\rm Top}^{\rm Bottom} {\rho \over \omega}
   d l.  \eqno(2.11) $$
The integration is along ${\vec \omega}$-lines. For closed vortex lines "Top"
is the
 same as "Bottom" and a "cut" must be introduced. These surfaces may be
parametrized
  $\lambda(\alpha)$.
The metage $\mu$, is defined
as in (2.11) taken up to some \lq\lq red mark" on the ${\vec \omega}$-line:
$$  \mu = \int_{\rm Bottom}^{\rm Red~mark} {\rho \over \omega} dl.
    \eqno(2.12) $$
 Notice that if the vortex lines are closed $\mu$ is an angular variable.
The variable  $\beta$ is defined by a family of surfaces of
$\vec \omega$-lines as well .  Thus, by definition,
$$ {\vec \omega} \cdot {\vec \nabla} \alpha = 0, \qquad
   \vec \omega \cdot  \vec \nabla \beta = 0 \qquad {\rm and} \qquad
\vec \omega \cdot \vec \nabla \mu = \rho.     \eqno(2.13) $$
The parametrization of $\alpha$ and $\beta$ may be so chosen that
 $$ {\vec \omega} = {\vec \nabla} \alpha \times {\vec \nabla} \beta.
    \eqno(2.14)$$
 Having chosen $\alpha^k = \alpha, \beta, \mu$,
It follows from (2.14) that
$$ \rho =
   {\partial (\alpha, \beta, \mu) \over \partial (x, y, z)}
   = \vec \nabla \alpha \times \vec \nabla \beta \cdot \vec \nabla \mu.
\eqno(2.15)$$
 We can translate $\mu$ with out changing $L$ if vortex lines are closed, or
$L$
 is not sensitive to translations in the $\mu$ direction.

The appropriate symmetry displacement associated with the infinitesimal change
in $\alpha$ is:
$$ \vec \xi = -{\partial \vec r \over \partial \alpha^k}\delta \alpha^k
\eqno(2.16) $$
Assuming that vortex lines are closed or translations of  $\mu$ does not effect
 $L$ for
 some reason [one may inquire the nature of flows of which a translation of
$\mu$
does not effect  $L$, a short calculation will show that the condition for this
is
Moffat's (1969) condition: $\int_S ({1 \over 2} \vec u^2 - h -\phi) \vec \omega
\cdot d \vec S  = 0$ as expected] we have the symmetry displacement:
$$ \vec \xi = -{\partial \vec r \over \partial \mu}\delta \mu \eqno(2.17)$$
Using equation (2.7)  we obtain the conservation laws:
$$ \int_V \vec u \cdot {\partial \vec r \over \partial \mu}  d^3\alpha = const.
  \eqno(2.18) $$
Now from (2.15) we see that:
$$ \rho {\partial \vec r \over \partial \mu} = \vec \omega \eqno(2.19)$$
and thus from (2.18) we obtain the helicity conservation law:
$$ {\cal H} = \int_{V} \vec u \cdot \vec \omega d^3x = const. \eqno(2.20)$$
Thus we conclude that the alpha
translation group in the direction of $\mu$ generates conservation of helicity.
[ One could ofcourse introduce the symmetry displacement $\vec \xi = \epsilon
{\vec \omega
\over \rho}$ right after equation (2.7), however, in this case one should show
that the above
displacement is a symmetry group displacement
which is not obvious if we do not take into account Arnold's
group and LBK's labeling. Moreover in coordinate space the symmetry group
appears aribitiraly
complex depending on the flow considered as opposed
to its apparent simplicity in alpha space.]
\vfill\eject

\noindent
{\bf References}

\refs

Arnold, V.I. 1966, {\it J. M\'ec.}, {\bf 5}, 19.

Bretherton, F.P. 1970 {\it J.  Fluid Mech}, {\bf 44}, 117.

Lynden-Bell, D. and Katz, J. 1981, {\it Proc. Roy. Soc. London}, A{\bf 378},
179 (referred to as LBK81).

Moffat H.K. 1969, {\it J. Fluid Mech}, {\bf 35}, 117.

Moreau, J.J. 1977, {\it Seminaire D'analyse Convexe, Montpellier 1977 , Expose
no: 7}

\endrefs
\vfill\eject

\bye